\documentclass{article}%
\usepackage{amsmath}%
\usepackage{amsfonts}%
\usepackage{amssymb}%
\usepackage{graphicx}
\usepackage[dvips,final]{epsfig}
\usepackage[latin1]{inputenc}

\makeatletter

\newenvironment{figurehere}
  {\def\@captype{figure}}
  {}
\makeatother

\begin{document}

\title{Cosmological Equation of State and Interacting Energies.}
\author{Alberto C. Balfagón(1) and Raúl Ramírez-Satorras(2).
\\Sección de Física Teórica y Aplicada. 
\\Departamento de Ingeniería Industrial,
\\Instituto Químico de Sarriá, Barcelona (Spain)
\\Vía Augusta 390, Barcelona, Spain
\\
\\(1)	albert.balfagon@iqs.edu
\\(2) raulramirezs@iqs.es}
\date{}
\maketitle

\begin{abstract}
In this paper we study a model of cosmic evolution, assuming that the different components of the universe could interact between them any time. An effective equation of state (EOS) for the universe has been used as well. A particular function for $w$, which gives a good agreement between our results and the experimental data, has been studied. Finally, the model obtained has been applied to different important cases.
\\
\begin{flushleft}
\textbf{Keywords:} dark energy theory, cosmological simulations, gravity.
\end{flushleft}
\end{abstract}

\tableofcontents

\section{Introduction}
At the end of the last century, it was experimentally showed that our universe has recently started an acceleration era [1, 2, 3]. One of the explanations to this phenomenon is based on the existence of a dark energy, which draw new attention to the cosmological constant, see [4,5] for a review and [6] for the study of the properties of dark energy from recent observations. This simple explanation had some problems that were faced using different models [7, 8, 9].

Most of the models assume that the components of the universe do not interact and look for an equation of state for the dark energy that could solve the problems [10], several authors imposed cosmological [11] or energetic [12] conditions in order to delimit the parameters.
Sometimes, the studied models allow the variation of the equation of state with time [13]; there are also models that consider the possibility of the evolution of the dark energy density along the time [14, 15, 16]. On the other hand, other authors directly change the Friedmann's equations [17]. 
Several authors investigated the possibility of interaction, but limited to some components of the universe, for instance [18, 19, 20, 21] consider the interaction between dark energy and dark matter.Recently some authors studied observational constraints for interactions between dark energy and dark matter [22,23,24,25]. Phantom cosmological models have also been proposed but, aside its physically odd negative energy, they seem not help to explain the problem correctly [26,27]. There are very few models that study the possibility of an effective EOS for the universe [13, 20], however they only work with some components and with red shift not very far from the present.

In this paper a general model for the universe has been studied, with very few restrictions. Thus the number of the constituents of the universe has been not fixed and interactions between any kinds of components in any time of the history of the universe have been permitted. The work has been carried out with an effective EOS for the universe that is taken as $p=w\rho$, where any variable can depend on $z$. The paper is divided in three sections; in the first one, the general equations that have been used are presented. In the next section a particular form of $w$ has been studied, fixing the different parameters in order to get the best agreement with the experimental data [1]. Finally, in the last section, the model obtained has been applied to some particular situations that let to some interesting deductions about the past and future cosmic evolution.

\section{General equations.}
The units used throughout this paper will be physical units in order to be more pedagogical and obtain equations more intuitive. As is currently done nowadays, the study has been carried out considering that our universe is flat. Friedmann's equations with the last assumptions are [28]: 

\begin{equation}
	\rho'=\frac{c^2}{8\pi G}3\left(\frac{\dot{a}}{a}\right)^2-\frac{c^4}{8\pi G}\Lambda
\end{equation}

Where $\rho'$ stands for any kind of energy but the one that comes from  $\Lambda$ (the so-called cosmological constant, however in this paper, the possibility that it could change with $z$ is been taken into account).

\begin{equation}
	p'=-\frac{c^2}{8\pi G}\left(2\frac{\ddot{a}}{a}+\left(\frac{\dot{a}}{a}\right)^2\right)+\frac{c^4}{8\pi G}\Lambda
\end{equation}

Where $p'$, as in Eq.(1), takes into account any pressure except for the one due to $\Lambda$. This pressure $p'$ can be thought as the pressure made by all the components of the universe (without the pressure from $\Lambda$ ) together with the pressure due to all kinds of interactions between all the constituents of the universe (considering now the interactions where $\Lambda$  appears).
Eq.(1) and (2) can now be rewritten as:

\begin{equation}
	\rho=\rho'+\frac{c^4}{8\pi G}\Lambda=\rho'+\rho_\Lambda=\frac{c^2}{8\pi G}3\left(\frac{\dot{a}}{a}\right)^2
\end{equation}

Where $\rho$ is the total energy density of the universe.

\begin{equation}
	p=p'-\frac{c^4}{8\pi G}\Lambda=p'+p_\Lambda=-\frac{c^2}{8\pi G}\left(2\frac{\ddot{a}}{a}+\left(\frac{\dot{a}}{a}\right)^2\right)
\end{equation}

Where $p$ is the total pressure of the universe.
From now on, all Friedmann's equations that will be used are Eq.(3) and (4), with the last interpretations for $\rho$ and $p$.

From Eq.(3) and (4) we can get:

\begin{equation}
	\dot{\rho}=-3\frac{\dot{a}}{a}\left(\rho+p\right)
\end{equation}

The cosmic acceleration can be deduced from  $\frac{\ddot{a}}{a}$:

\begin{equation}
	\frac{\ddot{a}}{a}=-\left(3p+\rho\right)\frac{4\pi G}{3c^2}
\end{equation}

The relation between the energy density and the pressure will be taken as:

\begin{equation}
	p=w\rho
\end{equation}

Where $w$ and $\rho$ can be functions of the red shift.
Applying Eq.(7) in (6) it can be deduced the necessary condition for the acceleration of the universe.

\begin{equation}
	w<-\frac{1}{3}
\end{equation}

Using (5) and (7) we get:

\begin{equation}
	\rho=\rho_0\exp\left(3\int^z_0 \frac{1+w}{1+z}dz\right)
\end{equation}

Where sub index 0 stands for the present time. 
The Hubble function, deduced from (3) and (9), is:

\begin{equation}
	H=\frac{\dot{a}}{a}=\left(\frac{8\pi G}{3c^2}\right)^{1/2}\left(\rho_0\exp\left(3\int^z_0 \frac{1+w}{1+z}dz\right)\right)^{1/2}
\end{equation}
 
The fitting of the different parameters used will be done through the experimental data from the SNeIa [1]. The experimental data are expressed by the distance modulus for different red shifts, and its relation with the Hubble function is [7,28]:  

\begin{equation}
	D_L=\left(1+z\right)c\int^z_0 \frac{dz}{H}
\end{equation}

Where $D_L$ is the light distance.
The distance modulus $\Delta$ is defined as the difference between the apparent and absolute magnitudes for each one of the SNeIa supernovas, its relation with the light distance is: 

\begin{equation}
	\Delta=25+5\log D_L
\end{equation}

Where $D_L$ is expressed in Mpc.

\section{Equation of state (EOS)}
In this section one of the many possible forms for the equation of state (7) will be studied. The reason for the selection done is similar to the followed by other authors [13]: $w$ must fit the EOS for radiation for the early universe, vary with time (or as in an equivalent manner, vary with the red shift) and allow that in the present time, the universe could attain an acceleration era. 
The $w$ will be:

\begin{equation}
	w=\frac{1}{3}\frac{\left(1+z\right)}{\left(\alpha+z\right)}-\gamma\frac{\left(\beta-1\right)}{\left(\beta+z\right)}
\end{equation}

Where $\alpha$, $\beta$ and $\gamma$ are three constants that will be fixed using the experimental data.
With Eq.(13) the Eq.(9), (10) and (12) can be expressed as:

\begin{equation}
	\rho=\rho_0\left(\frac{z+\beta}{\beta}\right)^{3\gamma}\frac{\left(\alpha+z\right)}{\alpha}\left(1+z\right)^{3\left(1-\gamma\right)}
\end{equation}

\begin{equation}
	H=H_0\left(\frac{z+\beta}{\beta}\right)^{3\gamma/2}\left(\frac{\alpha+z}{\alpha}\right)^{1/2}\left(1+z\right)^{\frac{3\left(1-\gamma\right)}{2}}
\end{equation}

\begin{equation}
	\Delta=25+5\log\frac{c\left(1+z\right)\left(\alpha\beta^{3\gamma}\right)^{1/2}}{H_0}\int^z_0\frac{1}{\left(z+\beta\right)^{\frac{3\gamma}{2}}\left(\alpha+z\right)^{\frac{1}{2}}\left(1+z\right)^{\frac{3\left(1-\gamma\right)}{2}}}dz
\end{equation}

The limits studied for each one of the parameters $\alpha$, $\beta$ and $\gamma$ have been:

	\[\alpha\in\left[3, 4.5\right], \beta\in\left[3, 4.5\right], \gamma\in\left[0.5, 1.5\right]
\]

These limits were fixed in order to let $w$ fit, as much as possible, the known history of the universe, mainly the recently accelerated era for $z$ between 1 and 2.
To study the best set of parameters $\alpha$, $\beta$ and $\gamma$ a full factorial design of experiments (DOE) [29] has been implemented, in order to obtain the best fitting between Eq.(16) and the 307 SNeIa distance modulus experimental data [1]
Taking ~$H_0=100h$ , where $h$ can be thought as a new free parameter to be fixed, the best set of parameters were:

\begin{equation}
	\alpha= 4.1, \beta= 3.5, \gamma= 1.0, h= 0.697
\end{equation}

With the set of parameters above mentioned (17) the mean absolute error found was 0.23, while the mean absolute error for a model that consider no interactions and only with baryonic matter $(w = 0)$, dark matter $(w = 0)$ and dark energy $(w = -1)$, was 0.23.
Eq.(13)-(16) can now be expressed, taking $\gamma= 1.0$, as:

\begin{equation}
	w=\frac{1}{3}\frac{\left(1+z\right)}{\left(\alpha+z\right)}-\frac{\left(\beta-1\right)}{\left(\beta+z\right)}
\end{equation}

\begin{equation}
	\rho=\rho_0\left(\frac{z+\beta}{\beta}\right)^3\frac{\left(\alpha+z\right)}{\alpha}
\end{equation}

\begin{equation}
	H=H_0\left(\frac{z+\beta}{\beta}\right)^{3/2}\left(\frac{\alpha+z}{\alpha}\right)^{1/2}
\end{equation}

\begin{equation}
	\Delta=25+5\log\left[2\frac{c\left(1+z\right)\left(\alpha\beta^3\right)^{1/2}}{H_0\left(\alpha-\beta\right)}\left(\sqrt{\frac{\alpha}{\beta}}-\sqrt{\frac{\alpha+z}{\beta+z}}\right)\right]
\end{equation}

In Fig.1 it can be seen the evolution of Eq.(18) where $z\in\left[-1,10\right]$. The red shift for the start of the acceleration (8) is inside $z\in\left[1.6,1.7\right]$. It can be deduced that the EOS of the universe will tend to a cosmological constant EOS: $w\rightarrow-1$, then $p\rightarrow-\rho$.

\begin{figurehere}
	\centering
	\vspace{0.5cm}
	\hspace{-3cm}
	\includegraphics[width=10cm]{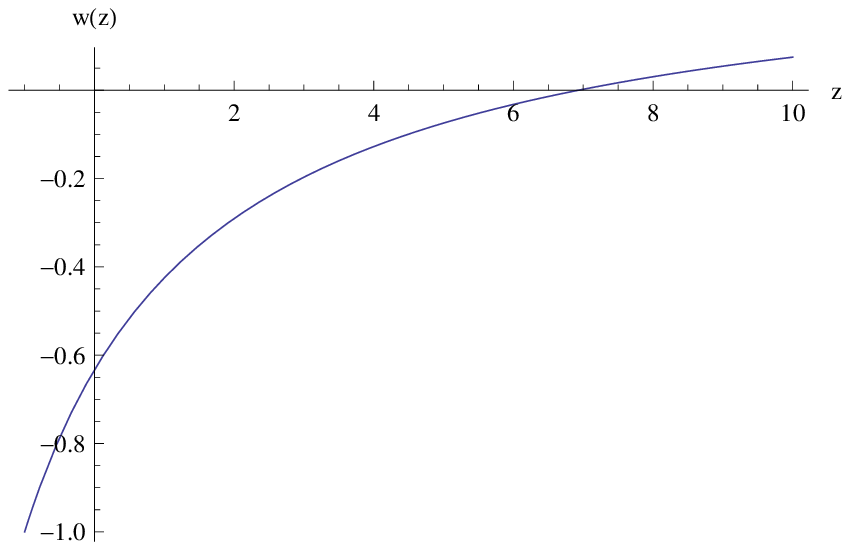}
	\caption{Function $w(z)$ against $z$, using constants(17)}
	\vspace{0.5cm}
	\label{fig:fig1}
\end{figurehere}
  
In Fig.2 the evolution of the total pressure (7) in function of $z$, within the interval $z\in\left[-1,8\right]$ it is shown.

\begin{figurehere}
	\centering
	\vspace{0.5cm}
	\hspace{-3cm}
	\includegraphics[width=10cm]{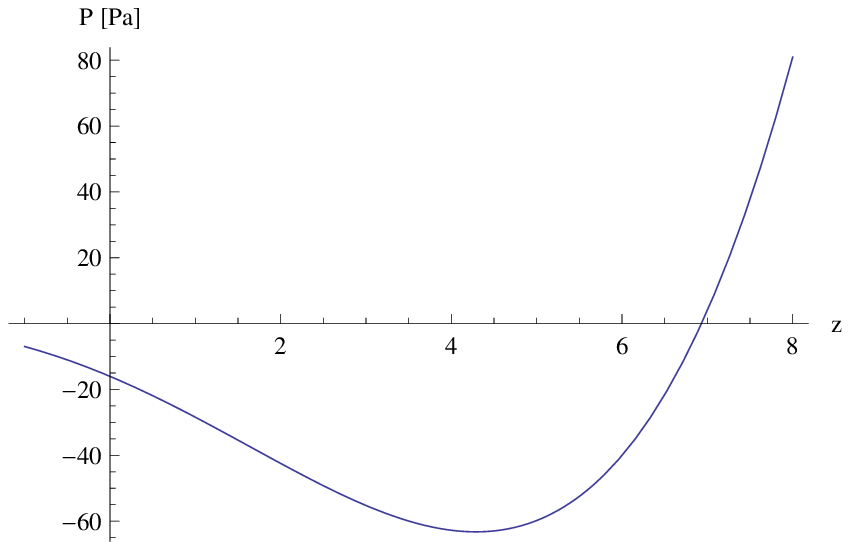}
	\caption{Function $p(z)$ against $z$, using constants (17)}
	\vspace{0.5cm}
	\label{fig:fig2}
\end{figurehere}

In Fig.2 it can be observed that the universe reached a total negative pressure era, before the present acceleration era, where gravitation still dominated during an interval of time. The continued expansion of the universe made possible that the negative pressure overcame gravitation driving to the accelerated cosmic expansion.

In Fig.3 the experimental values [1] are plotted against the obtained from Eq.(21) using the parameters in (17).
 
\begin{figurehere}
	\centering
	\vspace{0.5cm}
	\hspace{-3cm}
	\includegraphics[width=10cm]{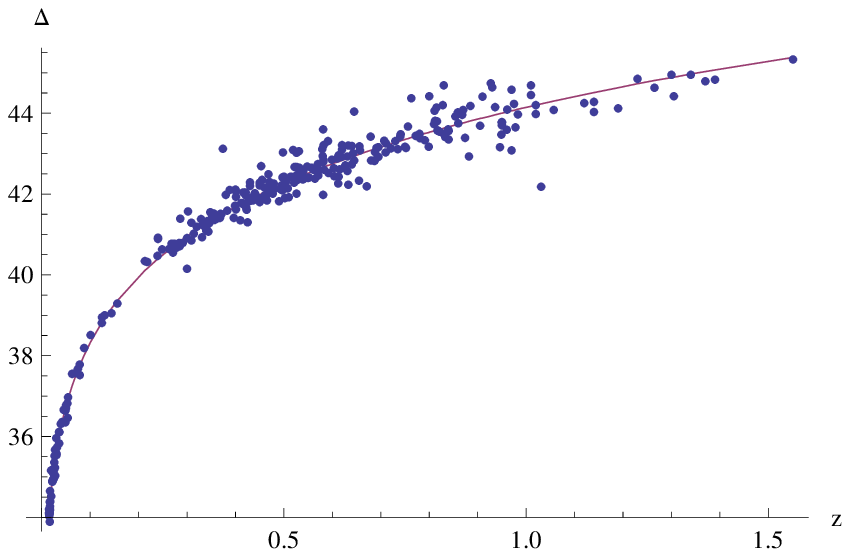}
  \caption{Distance modulus against $z$. Points are experimental data and the line represent the values calculated with (21) using (17).}
  \vspace{0.5cm}
	\label{fig:fig3}
\end{figurehere} 

Figs.4, 5 and 6 are magnifications of the graphs showed in Fig.3 for $z\in\left[0, 0.5\right]$ Fig.4, $z\in\left[0.5, 1\right]$ Fig.5 and $z\in\left[1,1.5\right]$ Fig. 6.

\begin{figurehere}
	\centering
	\vspace{0.5cm}
	\hspace{-3cm}
	\includegraphics[width=10cm]{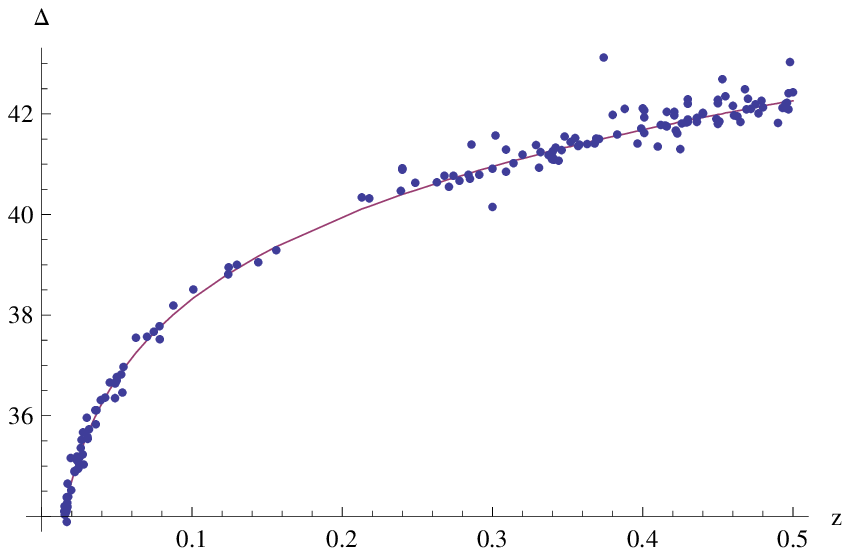}
  \caption{Magnification of Fig. 3 for $z$ between 0 and 0.5.}
  \vspace{0.5cm}
	\label{fig:fig4}
\end{figurehere}  

\begin{figurehere}
	\centering
	\vspace{0.5cm}
	\hspace{-3cm}
	\includegraphics[width=10cm]{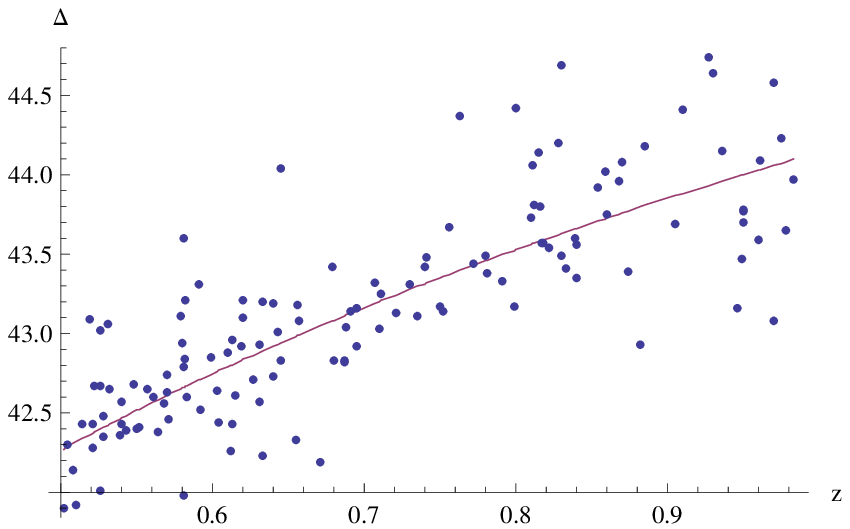}
  \caption{Magnification of Fig. 3 for $z$ between 0.5 and 1}
  \vspace{0.5cm}
	\label{fig:fig5}
\end{figurehere}  

\begin{figurehere}
	\centering
	\vspace{0.5cm}
	\hspace{-3cm}
	\includegraphics[width=10cm]{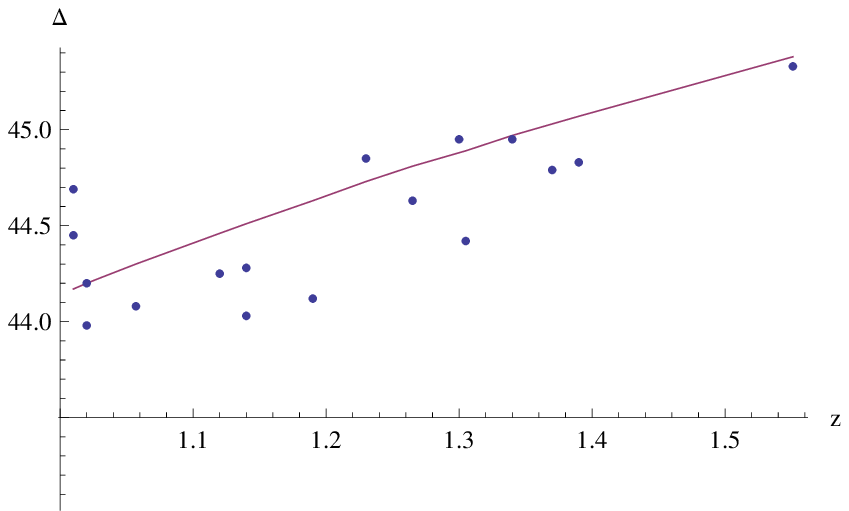}
  \caption{Magnification of Fig. 3 for $z$ between 1 and 1.5.}
  \vspace{0.5cm}
	\label{fig:fig6}
\end{figurehere}  

An important issue is to know, as much as possible, the evolution of the cosmic acceleration. Using Eq.(6), and with some sensible assumptions, it is possible to get some useful information about the evolution of the cosmic acceleration.
Assuming that the scale factor $a$ is an ever growing function (reasonable assumption with the present experimental data that we have), if we express Eq.(6), using (7), as:

	\[\frac{\ddot{a}}{a}=-\left(3w+1\right)\rho\frac{4\pi G}{3c^2}
\]

then if  $-(3w+1)\rho$ raises this will imply that the cosmic acceleration will raise, if it diminishes we cannot obtain any conclusive conclusion because what could happen is  that the acceleration diminishes or raises more slowly than the scale factor $a$. In Fig.7 we have plotted $-(3w+1)\rho$ in function of $z$.

\begin{figurehere}
	\centering
	\vspace{0.5cm}
	\hspace{-3cm}
	\includegraphics[width=10cm]{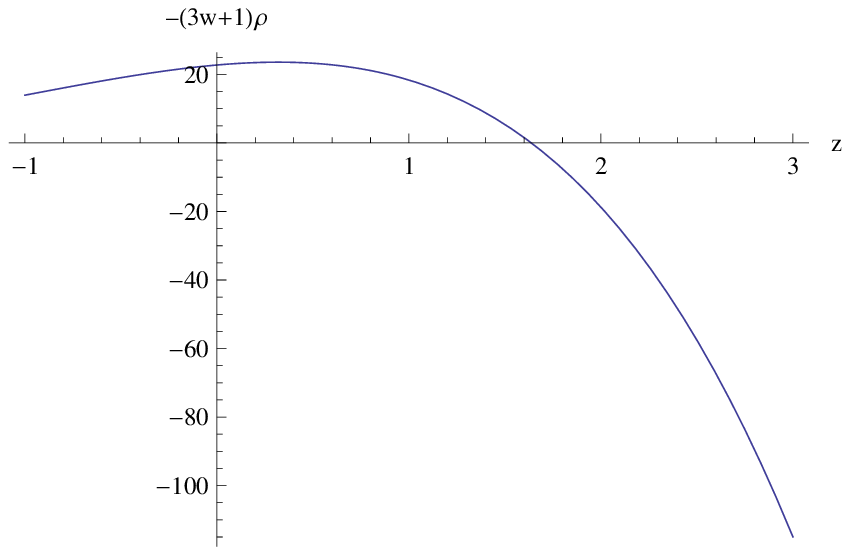}
  \caption{ Representation of  $-(3w+1)\rho$ in function of $z$.}
  \vspace{0.5cm}
	\label{fig:fig7}
\end{figurehere}

From the data (as it can be seen in Fig.7) it can be deduced that the cosmic acceleration rose until $z\in\left[0.31,0.33\right]$ and from this point forward it is not possible to assure the exact acceleration's evolution. Of course, it can also be seen that at the point when the cosmic acceleration is positive it will remain positive forever.

\section{Particular cases.}

In this section, three models have been studied applying the equations that have been deduced in the previous section.

\subsection{Model with interaction between the components of the universe and constant values for the dark energy density and pressure.}

In this model Eq.(7) is taken as the sum of the next terms:

\begin{equation}
	p=w\rho=p_m+p_r+p_\Lambda +p_i
\end{equation}

Where $p_m$ is the baryonic and dark matter pressure,  $p_r$ is the radiation pressure, $p_\Lambda$ is the dark energy pressure and $p_i$ is a pressure that comes out from the interaction of the components of the universe.
The EOS for each one of the components will be taken as if there was no interaction between them, so loading any kind of interaction in $p_i$. The pressures for the different components are:

Baryonic and dark matter pressure:  $p_m=0$

Dark energy pressure:  
\begin{equation}
	p_\Lambda=-\rho_\Lambda=-\frac{c^4\Lambda}{8\pi G}
\end{equation}

Radiation pressure:
\begin{equation}
	p_r=\frac{1}{3}\rho_r=\frac{1}{3}\rho_{ro}\left(1+z\right)^4
\end{equation}

From the currently accepted adimensional densities values [28] $\Omega_\Lambda^0=0.7$ and $\Omega_r^0=0.0001$ the expressions (23) and (24) result on:

\begin{equation}
	p_\Lambda=-0.7\frac{3c^2H^2_0}{8\pi G}
\end{equation}

\begin{equation}
	p_r=\frac{0.0001}{3}\frac{3c^2H^2_0}{8\pi G}\left(1+z\right)^4
\end{equation}

The interaction pressure can be deduced substituting (18), (19), (25) and (26) in (22):

	\[p_i= \frac{3c^2H^2_0}{8\pi G} \left(\frac{z+\beta}{\beta}\right)^3\frac{\left(\alpha+z\right)}{\alpha}\left(\frac{1}{3}\frac{\left(1+z\right)}{\left(\alpha+z\right)}-\frac{\left(\beta-1\right)}{\left(\beta+z\right)}\right)+
\]
\begin{equation}
	\frac{3c^2H^2_0}{8\pi G}\left(0.7 -\frac{0.0001}{3}\left(1+z\right)^4\right)
\end{equation}

In Fig.8 the interaction pressure for $z\in\left[-1,8\right]$ is shown.

\begin{figurehere}
	\centering
	\vspace{0.5cm}
	\hspace{-3cm}
	\includegraphics[width=10cm]{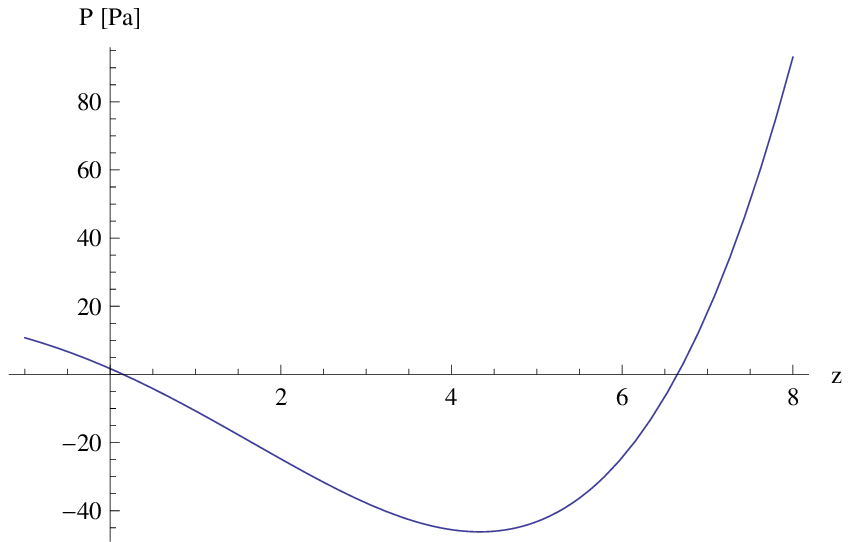}
  \caption{Interaction pressure as a function of $z$}
  \vspace{0.5cm}
	\label{fig:fig8}
\end{figurehere}  
  
From the figure above, different interpretations can be drawn; one of these, could be that the interactions between the different components of the universe gave a positive pressure while the baryonic matter and radiation densities were large. With the expansion of the universe and the decrease of densities, an interaction between dark energy and the baryonic matter started to dominate (so dark energy would not be as dark as it is assumed to be), which resulted in a negative interaction pressure. Eventually, when the baryonic matter density was very small, an initially small interaction between dark energy and dark matter started to dominate, which resulted in a positive interaction pressure. 

Thus, some experimental sign of an interaction between dark energy and baryonic matter could be a signal that the above interpretation could be possible.

Other possible interpretation could be the possibility of a self-interacting dark matter [30] that could provide an accelerated expansion phase as it was shown in [31, 32]

\subsection{Model without interaction and with a variable dark energy pressure.}

In this model Eq.(22) can be put as:

\begin{equation}
	p=w\rho=p_m+p_r+p_\Lambda
\end{equation}

The dark energy pressure can be expressed in a general form as:

\begin{equation}
	p_\Lambda=w_\Lambda\rho_\Lambda
\end{equation}

Working in a similar way as it was done to get Eq(27) the following result is obtained: 

	\[p_\Lambda=\frac{3c^2H^2_0}{8\pi G}\left(\frac{z+\beta}{\beta}\right)^3\left(\frac{\alpha+z}{\alpha}\right)\left(\frac{1}{3}\frac{\left(1+z\right)}{\left(\alpha+z\right)}-\frac{\beta-1}{\beta+z}\right)-
\]

\begin{equation}
	\frac{3c^2H^2_0}{8\pi G}\left(\frac{0.0001}{3}\left(1+z\right)^4\right)
\end{equation}

At this point, different ways can be followed depending on which term in (29) it is considered not constant.

\subsubsection{$\rho_\Lambda=cte$}

From equations (29) and (30), and taking $\Omega_\Lambda^0=0.7$, it can be deduced that  $w_\Lambda$ will be expressed as:

\begin{equation}
	w_\Lambda=\frac{1}{0.7}\left[\left(\frac{z+\beta}{\beta}\right)^3\left(\frac{\alpha+z}{\alpha}\right)\left(\frac{1}{3}\frac{\left(1+z\right)}{\left(\alpha+z\right)}-\frac{\beta-1}{\beta+z}\right)-\frac{0.0001}{3}\left(1+z\right)^4\right]
\end{equation}

In Fig.9 $w_\Lambda$ has been plotted, Eq.(31), for $z\in\left[-1,8\right]$

\begin{figurehere}
	\centering
	\vspace{0.5cm}
	\hspace{-3cm}
	\includegraphics[width=10cm]{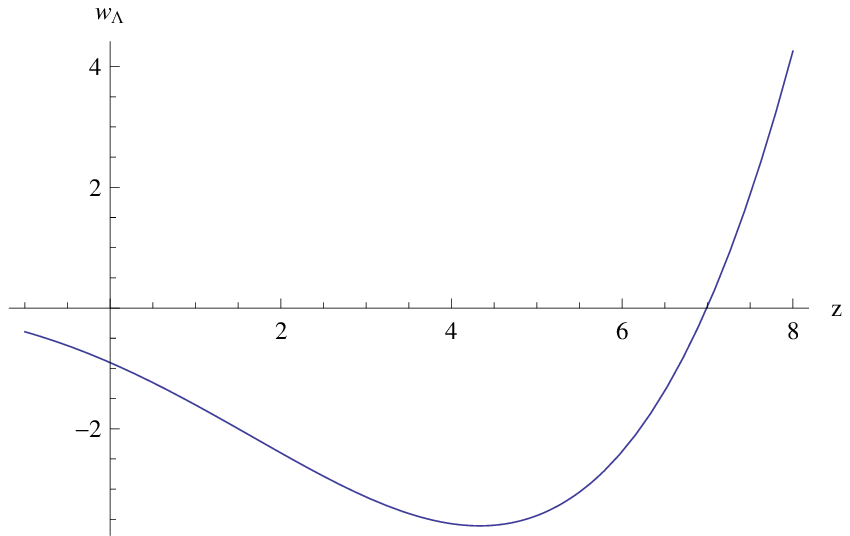}
 	\caption{$w_\Lambda$ as a function of $z$.}
  \vspace{0.5cm}
	\label{fig:fig9}
\end{figurehere}  

Comparing Fig.9 with Fig.1 it can be observed that while the $w$ of the universe tends to -1, the dark energy $w_\Lambda$ tend to -0.4 and attain values lower than -1 during some time. This difference in behavior appears due to the fact that in Fig.1 $w$ and the energy density of the universe are free to be time (or z) functions, while in Fig. 9 a constant dark energy density is fixed. This result is interesting because it shows that when some authors claim for the necessity of a phantom period for the EOS of the dark energy, in fact it is not necessary because what happen are a variation of the energy density and the $w$ of the universe, keeping always $w\geq-1$.
In the same way, the results obtained show that it is very difficult to consider that the dark energy density is constant (it leads to a phantom energy), so it must change during the cosmic evolution.

\subsubsection{$\rho_\Lambda\neq cte$}

In this case the EOS for the dark energy density will be:

\begin{equation}
	p_\Lambda=-\rho_\Lambda
\end{equation}

With (32) the Eq.(28) let the determination of the density parameter $\Omega_\Lambda$ (which nowadays is considered to be $\Omega^0_\Lambda=0.7$) as a function of $z$:

\begin{equation}
	\Omega_\Lambda=\frac{0.0001\left(1+z\right)^4}{3}-\left(\frac{z+\beta}{\beta}\right)^3\left(\frac{\alpha+z}{\alpha}\right)\left(\frac{1}{3}\frac{\left(1+z\right)}{\left(\alpha+z\right)}-\frac{\beta-1}{\beta+z}\right)
\end{equation}

In Fig. 10 the Eq.(33) for $z\in\left[-1,10\right]$ is represented. It can be observed that the density parameter took negative values in the far past and nowadays $\left(z=0\right)$ tends to 0.63.

\begin{figurehere}
	\centering
	\vspace{0.5cm}
	\hspace{-3cm}
	\includegraphics[width=10cm]{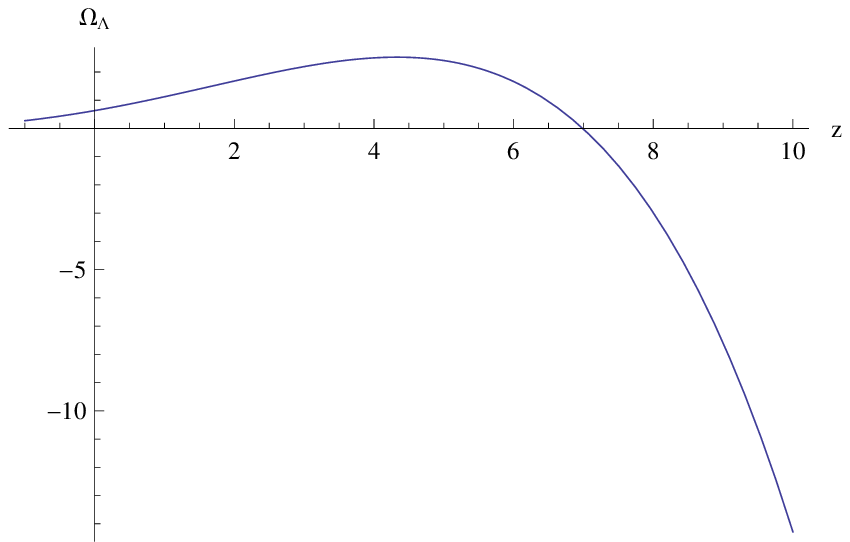}
 	\caption{Parameter density for $z$ running between -1 and 10.}
  \vspace{0.5cm}
	\label{fig:fig10}
\end{figurehere}  

A negative energy is not very acceptable, so model 4.2.2 is not physically correct. However, given the fact that for $z=0$ a value similar to the accepted one is obtained, it can be deduced that if the dark energy density varies with $z$ then this variation must be similar to the variation showed in Eq.(33). In order to avoid negative values for the density it has to be assumed that $w_\Lambda$ can vary and/or that there must have an interaction pressure. 

\subsection{Model with interaction and a variable dark energy pressure.}

In this last case a model with interaction between the different components of the universe (expressed as an interaction pressure) is studied. We will also admit that the dark energy density could vary, assuming an EOS for dark energy of the cosmological constant type: $p_\Lambda=-\rho_\Lambda$.
The variation of the dark energy density will be deduced from Eq.(1), expressed now using the Hubble function $H$ as:

\begin{equation}
	H^2=\frac{8\pi G}{3c^2}\rho'+\frac{c^2}{3}\Lambda
\end{equation}

Where $\Lambda$ is considered as a function that could change with $z$. The most general variation could be, from Eq.(34), expressed as:

\begin{equation}
	\Lambda=\Phi H^2 +\Psi \rho'+\Lambda_0
\end{equation}

Where $\Phi,\Psi$ and $\Lambda_0$ are considered constants. It is easy to show that the general expression (35) is equivalent to: 

\begin{equation}
	\Lambda=\alpha_1\rho'+\alpha_0
\end{equation}

Where $\alpha_0$ and $\alpha_1$ are new constants that are function of the previous ones.
Taking into account all what has been stated, the dark energy density can be expressed as:

\begin{equation}
	\rho_\Lambda=\frac{c^4}{8\pi G}\Lambda=\beta_1\rho'+\beta_0
\end{equation}

Where $\beta_0$ and $\beta_1$ are new constants to be fixed.
Considering the present values for the dark energy and matter (baryonic and dark) density parameters it can be deduced that $\alpha_0=0$ and $\alpha_1=\frac{56\pi G}{3c^4}$, so Eq.(36) can be written as (in agreement with [13]):

\begin{equation}
	\Lambda=\frac{56\pi G}{3c^4}\rho'
\end{equation}

And Eq.(37) will be:

\begin{equation}
	\rho_\Lambda=\beta_1\rho'=\frac{7}{3}\rho'
\end{equation}

Taking $\rho=\rho'+\rho_\Lambda$ and with (39) it can be deduced:

\begin{equation}
	\rho_\Lambda=\frac{\beta_1}{1+\beta_1}\rho=\frac{7}{10}\rho
\end{equation}

Eq.(40), and using (19), gives the same $z$ dependence for the dark energy density parameter as the one it was deduced in case 4.2.2. The constant $\beta_1$ could be let as a free parameter but we have taken the value deduced from (38) and (39) to use later. 

In this case the Eq.(7) will be the sum of the next pressures:

\begin{equation}
	p=w\rho=p_m+p_r+p_\Lambda+p_i
\end{equation}

Considering (40) and that the matter's pressure is zero, it can be deduced:

\begin{equation}
	\left(w+0.7\right)\rho=p_r+p_i
\end{equation}

Using (18), (19) and (26) in (42) we can deduce the interaction pressure:

	\[p_i=\frac{3c^2H^2_0}{8\pi G}\left(\frac{z+\beta}{\beta}\right)^3\frac{\left(\alpha+z\right)}{\alpha}\left(\frac{1}{3}\frac{\left(1+z\right)}{\left(\alpha+z\right)}-\frac{\beta-1}{\beta+z}+0.7\right)-
\]

\begin{equation}
	\frac{3c^2H^2_0}{8\pi G}\left(\frac{0.0001}{3}\left(1+z\right)^4\right)
\end{equation}

Fig.11 shows the interaction pressure which only reaches negative values on the future, a different behavior that showed in Fig.8.

\begin{figurehere}
	\centering
	\vspace{0.5cm}
	\hspace{-3cm}
	\includegraphics[width=10cm]{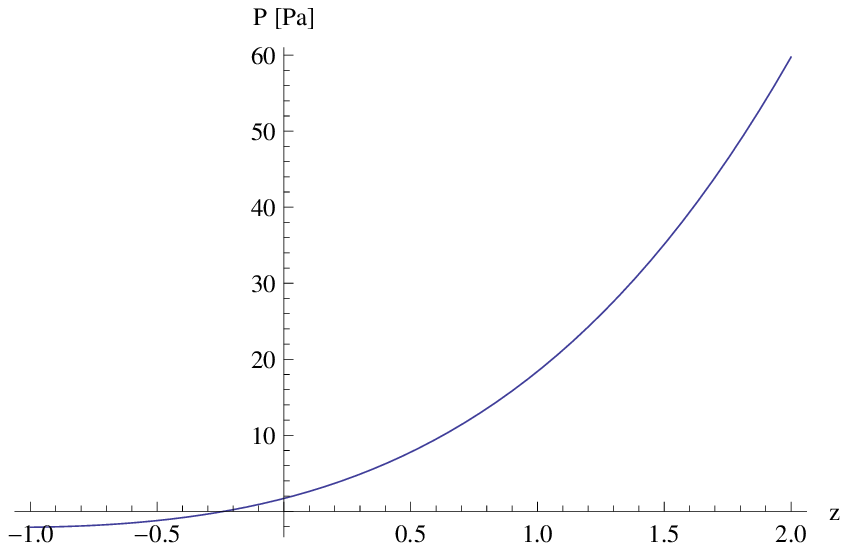}
	\caption{Interaction pressure for $z$ running from -1 to 2.}
  \vspace{0.5cm}
	\label{fig:fig11}
\end{figurehere}  

Finally Fig.12 shows the dark energy density (40), which always takes positive values.

\begin{figurehere}
	\centering
	\hspace{-3cm}
	\includegraphics[width=10cm]{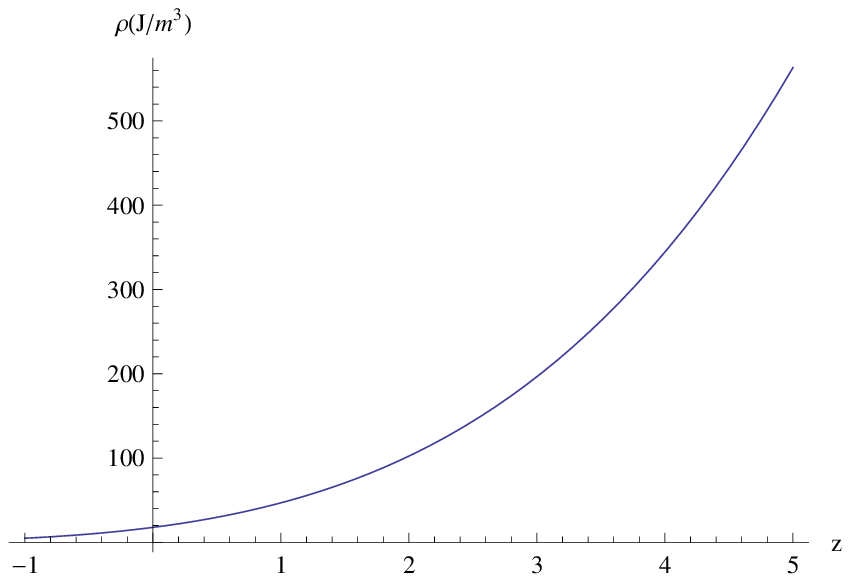}
	\caption{Dark energy density for $z$ running from -1 to 5.}
	\label{fig:fig12}
\end{figurehere}  

\section{Conclusions.}

In this paper it has been shown how using a model for the universe, where the interaction between their components is admitted, it is possible to fit an effective EOS with a result similar to the models that not consider interaction (so assuming that the total pressure is only the sum of the pressures of each component). The studied EOS for the fluid of the universe varies with time and has three parameters that have been calculated using the SNeIA experimental data.  The final effective EOS for the universe fits the EOS for radiation for the early universe and at present time allows an accelerated era.  

This general model provides several signatures that could be used to see its suitability. Taking $H_0=100h$ for the Hubble constant, the best value found for $h$ has been $0.697$. The predicted red shift, for the start of the actual phase of acceleration, is $z\in\left[1.6,1.7\right]$. From Fig. 7 it is noticed that the model also predicts a maximum for $\frac{\ddot{a}}{a}$ for a red shift $z\in\left[0.31,0.33\right]$. 

From the general model it is possible consider different sub-models, but each one will always fulfill the above properties. In the paper it has been considered three sub-models that are physically relevant: two models with interaction, one for constant dark energy density and pressure and the other for variable dark energy density and pressure. The third model studied was without interaction and variable dark energy pressure, including two sub-cases. Some interesting conclusions has been deduced from the above models: the lack of necessity to consider phantom energy in order to fit a suitable model for the SNeIa experimental data; On the other hand, the necessity of interaction between the different components of the universe to get positive dark energy density. Finally it has also been shown that there could be some kind of interaction in the dark sector (between dark energy and dark matter) and also between the dark sector and the baryonic matter.


\begin{thebibliography}{9}                                                                                                %

\bibitem {} M. Kowalski et al.,  Astrophys.J.686:749-778 (2008)
\bibitem {} L. Perivolaropoulos,  [astro-ph/0601014v2]
\bibitem {} T. Padmanabhan and T. Roy Choudhury, Mon. Not. Roy. Astron. Soc. 344:823-834 (2003)
\bibitem {} Ruth Durrer and Roy Maartens, Gen. Relativ. Gravit. (2008) 40: 301-328
\bibitem {} Raphael Bousso, Gen. Relativ. Gravit. (2008) 40: 607-637
\bibitem {} Puxun Wu and Hongwei Yu, JCAP0710:014,2007
\bibitem {} Steven Weinberg, Cosmology, Oxford University Press (2008).
\bibitem {} Michael S. Turner and Dragan Huterer, J.Phys.Soc.Jap.76:111015 (2007)
\bibitem {} A. Sil and S. Som, Astrophys.SpaceSci.318:109-115,2008
\bibitem {} Salvatore Capozziello, [gr-qc/0812.0170v1]
\bibitem {} Vinod B. Johri and P.K. Rath, Int.J.Mod.Phys. D16:1581-1591 (2007)
\bibitem {} M.P.Lima, S.D.P. Vitenti and M.J. Rebouças, Phys.Lett. B668 (2008) 83
\bibitem {} S.  Som and A.Sil, Astrophys. SpaceSci. 318:109-115 (2008)
\bibitem {} Arbab I Arbab, Class. Quantum Grav. 20 (2003) 93-99
\bibitem {} Utpal Mukhopadhyay et al., [gr-qc/0711.0686v1]
\bibitem {} A.A. Usmani et al., Mon. Not. Roy. Astron. Soc. Lett. 386:L92-95 (2008)
\bibitem {} Marek Szydlowski  et al., Phys.Lett. B642 (2006) 171-178
\bibitem {} Bo Feng, Xiulian Wang and Xinmin Zhang, Phys.Lett. B607 (2005) 35-41 
\bibitem {} Gabriela Caldera-Cabral , Roy Maartens and Arturo Ureña-López, [gr-qc/0812.1827v1]
\bibitem {} Winfried Zimdahl, Int.J.Mod.Phys. D14 (2005) 2319-2326
\bibitem {} Luis P. Chimento, Monica Forte and Gilberto M. Kremer, Gen.Rel.Grav.41:1125-1137,2009
\bibitem {} Chang Feng, Bin Wang, Elcio Abdalla and Ru-Keng Su, Phys.Lett.B665:111-119,2008
\bibitem {} E. Abdalla, L. R. Abramo, L. Sodre and B. Wang, Physics Letters B 673: 107-110 (2009)
\bibitem {} Bin Wang, Jiadong Zang, Chi-Yong Lin, Elcio Abdalla and S. Micheletti, Nucl.Phys.B778:69-84,2007
\bibitem {} Jian-Hua He and Bin Wang, JCAP 0806:010,2008 
\bibitem {} Xi-ming Chen, Yungui Gong and Emmanuel N. Saridakis, JCAP 0904:001,2009
\bibitem {} Genly Leon and Emmanuel N. Saridakis,[gr-qc/0904.1577]
\bibitem {} Jordi Cepa, Cosmología Física, Ediciones Akal (2007)
\bibitem {} Montgomery, Diseño y análisis de experimentos, Limusa Wiley (2002)
\bibitem {} Spergel  D.N. and Steinhardt P.J., Phys. Rev. Lett., 84,3760 (2000)
\bibitem {} Zimdahl W., Schwarz D.J., Balakin A.B. and Pavon D.,Phys. Rev. D., 64, 3501 (2001)
\bibitem {} Lima J.A.S., Silva F.E. and Santos R.C.,Class. Quantum Grav, 25, 205006 (2008)

\end{thebibliography}
\end{document}